\documentclass[twocolumn]{aastex62}
\usepackage{graphicx}
\usepackage{subfigure}
\usepackage{multirow}
\usepackage{enumerate}

\submitjournal{ApJ}

\def\ie{{\it i.e. }}
\def\eg{{\it e.g. }}

\def\vk{{\vec{k}}}
\def\kx2y2{{k_x^2-k_y^2}}
\def\wG{{\widetilde{G}}}
\def\hG{{\hat{G}}}

\shorttitle{Mitigating the Noise-Source Coupling Effect in Shear Measurement}
\shortauthors{Hekun Li et al.}

\begin{document}

\title{Mitigating the Noise-Source Coupling Effect in Shear Measurement}

\correspondingauthor{Jun Zhang}
\email{betajzhang@sjtu.edu.cn}

\author{Hekun Li}
\affiliation{Department of Astronomy, Shanghai Jiao Tong University, Shanghai 200240, China}
\author{Jun Zhang*}
\affiliation{Department of Astronomy, Shanghai Jiao Tong University, Shanghai 200240, China}
\affiliation{Shanghai Key Laboratory for Particle Physics and Cosmology, Shanghai 200240, China}

\begin{abstract} 
Sub-percent level accuracy in shear measurement is required by the Stage-IV weak lensing surveys. One important challenge is about suppressing the shear bias on source images of low signal-to-noise ratios (SNR$\lesssim10$). Previously, it has been demonstrated that the shear estimators defined in the Fourier\_Quad (FQ) method can achieve sub-percent accuracy at the very faint end (SNR$\lesssim5$) through ensemble averaging. Later, it is found that we can approach the minimum statistical error (the Cramer-Rao Bound) by symmetrizing the full PDF of the FQ shear estimators (the PDF\_SYM approach), instead of taking ensemble averages. Recently, with a large amount of mock galaxy images, we are able to identify some small amount of shear biases in the PDF\_SYM approach at the faint end. The multiplicative bias goes up to $1-2\times10^{-2}$ at SNR $\lesssim 10$, and the anisotropy of the point spread function (PSF) causes an additive bias that can reach a few times $10^{-4}$. We find that these biases originate from the noise-source coupling in the galaxy power spectrum. It turns out that this problem can be largely fixed by adding additional terms to the FQ shear estimators. The resulting multiplicative and additive biases can be significantly suppressed to the level of $10^{-3}$ and $10^{-5}$ respectively. These corrections substantially extend the available SNR range for accurate shear measurement with the PDF\_SYM approach.

\end{abstract}


\section{Introduction} \label{sec:intro}

The large scale structure in the dark matter field perturbs the light passing through and introduces a slight but coherent distortion to the observed shapes of background galaxies. This effect is called weak gravitational lensing or cosmic shear. The statistical analysis of weak gravitational lensing is one of the most promising probes for the large scale mass distribution and the expansion history of the universe \citep{Bartelmann2001, Hoekastra2008, Kilbringer2015}. The ongoing weak lensing surveys, such as DES \citep{Troxel2018}, KIDS \citep{Hildebrandt2016}, and HSC \citep{Hikage2019}, and the next generation surveys, such as LSST \citep{Abell2009} and Euclid \citep{Laureijs2011}, are accumulating an unprecedented amount of galaxy images for the improvement of shear measurement accuracy to the sub-percent level, and therefore more stringent constraint on the cosmological parameters. 

The shear measurement is currently subject to many systematic errors, such as model bias \citep{Bernstein2010,Voigt2010,Kacprzak2014}, noise bias \citep{Refregier2012,Kacprzak2014}, and selection bias \citep{Hirata2003}. In the model-fitting methods \citep{Miller2013, Zuntz2013}, if the sources could not be exactly modelled by the profile, the measurement would be biased, \ie model bias. If the response of a method to noise is nonlinear, it would give rise to the noise bias \citep{Refregier2012, Viola2014, Berstein2016}. \cite{Refregier2012} find that the noise bias becomes comparable with the shear signal at low SNR ($\lesssim10$). A selection on the combination of selection criteria, \eg magnitude, and SNR, is the common method to avoid the systematic bias at the faint end. An inappropriate selection would lead to the selection bias \citep{Hirata2003,Kitching2008, Miller2013, Fenech2017,Liu2018,Li2020}. 

We focus on the Fourier\_Quad method\citep{Zhang2008, Zhang2015} in this paper to investigate the effect of noise on shear measurement. The Fourier\_Quad method is a moment-based method that measures the shape from the Fourier transformation of the galaxy image without any assumption of the galaxy morphology. It has been demonstrated that the Fourier\_Quad method can achieve sub-percent accuracy at SNR $\lesssim 5$ by taking the ensemble averages of the shear estimators \citep{Zhang2015}. Nevertheless, \cite{Zhang2017} find that taking the ensemble average of the FQ estimators is not the optimal statistical approach. Instead, they propose to recover the shear signal by symmetrizing the full PDF of the FQ shear estimators (the PDF\_SYM approach). It is found that the new approach indeed allows us to achieve the optimal statistical error (Cramer-Rao Bound) in shear measurement. Recently, we find that the PDF\_SYM approach would be slightly biased by the faint sources. The typical magnitude of the multiplicative bias can reach $1-2\times10^{-2}$ at SNR $\lesssim 10$, and the additive bias is typically a few times $10^{-4}$ depending on the ellipticity of the PSF. It is the purpose of this paper to address the origin and solution of this problem.

In Section \ref{sec:FQ}, we briefly introduce the FQ shear estimators and the PDF\_SYM approach for shear recovery. We discuss the origin of the bias in Section \ref{sec:bias_cal}, and present a set of slightly modified FQ shear estimators as a solution to this problem in Section \ref{sec:bias_corr}. We conclude in Section \ref{sec:conclusion}.

\section{The Fourier\_Quad method}\label{sec:FQ}
The Fourier\_Quad method is a model-independent shear measurement method. Its shear estimators are defined in Fourier space as \citep{Zhang2017}:
\begin{eqnarray}
\label{shear_estimator}
G_1&=&-\int d^2\vec{k}(k_x^2-k_y^2)T(\vec{k})M(\vec{k})\\ \nonumber
G_2&=&-2\int d^2\vec{k}k_xk_yT(\vec{k})M(\vec{k})\\ \nonumber
N&=&2\int d^2\vec{k}\left[k^2-\frac{\beta^2}{2}k^4\right]T(\vec{k})M(\vec{k}), \\ \nonumber
U &=& -\beta^2\int d^2\vec{k}(k_x^4 - 6k_x^2k_y^2 + k_y^4)T(\vec{k})M(\vec{k}), \\ \nonumber
V &=& -4\beta^2\int d^2\vec{k}(k_x^3k_y - k_xk_y^3)T(\vec{k})M(\vec{k}).
\end{eqnarray}
where $M(\vec{k})$ is the power spectrum of the source image subtracting the contribution of the background and Poisson noise, \ie:
\begin{eqnarray}
\label{FQ_TM}
&&M(\vec{k})=\left\vert\widetilde{f}_I(\vec{k})\right\vert^2-F_I-\left\vert\widetilde{f}_B(\vec{k})\right\vert^2+F_B,\\ \nonumber
&&F_I=\frac{\int_{\vert\vec{k}\vert > k_c} d^2\vec{k}\left\vert\widetilde{f}_I(\vec{k})\right\vert^2}{\int_{\vert\vec{k}\vert > k_c} d^2\vec{k}}, \;\;\; F_B=\frac{\int_{\vert\vec{k}\vert > k_c} d^2\vec{k}\left\vert\widetilde{f}_B(\vec{k})\right\vert^2}{\int_{\vert\vec{k}\vert > k_c} d^2\vec{k}},
\end{eqnarray}
where $\widetilde{f}_I(\vec{k})$ and $\widetilde{f}_B(\vec{k})$ are the Fourier transformations of the source image and a neighboring image of background noise respectively. $F_B$ and $F_I$ are the estimates of the Poisson noise power spectrum (independent of $\vk$) of the background noise and the source image respectively. A critical wavelength, $k_c$, is required to avoid the contamination from the source power \citep{Zhang2015}. The factor $T(\vec{k})$ is the ratio of the power spectrum of the target isotropic Gaussian PSF to that of the original PSF, \ie$T(\vec{k}) = |\widetilde{W}_{\beta}(\vec{k})|^2/|\widetilde{W}_{P}(\vec{k})|^2$.
It is designed to convert the original PSF, $W_{P}(\vec{x})$, to an isotropic Gaussian form $W_{\beta}(\vec{x})$, so that the effect of PSF can be corrected rigorously and model-independently.
The scale radius of $W_{\beta}(\vec{x})$ is $\beta$, which should be slightly larger than that of $W_{P}(\vec{x})$ to avoid the singularities in the factor $T(\vec{k})$. 

Fourier\_Quad method provides two statistical approaches for shear recovery from the estimators. One way is to take the ensemble averages of the shear estimators directly:
\begin{eqnarray}
\label{method:mean}
\frac{\left\langle  G_1\right\rangle }{\left\langle  N\right\rangle }=g_1+O(g_{1,2}^3),\;\;\;\frac{\left\langle  G_2\right\rangle }{\left\langle  N\right\rangle }=g_2+O(g_{1,2}^3).
\end{eqnarray}
It has been shown that it is accurate to the second order in shear \citep{Zhang2015}. An appealing feature of Fourier\_Quad is that the inclusion of extremely faint sources or even point sources to the sample does not bias the shear measurement in Eq.(\ref{method:mean}). Nevertheless, this averaging method defined in Eq.(\ref{method:mean}) is not statistically optimal. In \cite{Zhang2017}, a new approach called PDF\_SYM is proposed for the FQ shear estimators, aiming at achieving the minimum statistical error (the Cramer-Rao Bound) without introducing systematic biases. The new approach recovers the shear signal by symmetrizing the full probability distribution function (PDF) of the shear-corrected estimators $\hat{G}_{1/2}$, 
\begin{eqnarray}
\label{method:pdf}
\hat{G}_{1/2} = G_{1/2} - \hat{g}_{1/2}(N \pm U),
\end{eqnarray}
where $\hat{g}_{1/2}$ is the assumed value of the shear signal. The $\hat{g}_{1/2}(N \pm U)$ term corrects the aniosotropic part caused by shear in $G_{1/2}$. \cite{Zhang2017} shows that the PDF of $\hat{G}_{1/2}$ would be maximally symmetric with respect to zero if $G_{1/2}$ is corrected by the true shear value, \ie $\hat{g}_{1/2}=g_{1/2}$. Note that the $V$ term defined in Eq.(\ref{shear_estimator}) is useful for the transformation of $U$ in the case of coordinate rotation.

The PDF\_SYM approach is a promising method for shear recovery in the framework of Fourier\_Quad. However, its accuracy has not been strictly tested at the very faint end. It is the purpose of this work to investigate the performance of the PDF\_SYM approach for sources of different SNR, and to propose necessary modifications.


\section{Bias due to the noise-source coupling}\label{sec:bias_cal}

\begin{figure*}[htbp]
	\centering
	\includegraphics[width=0.9\linewidth]{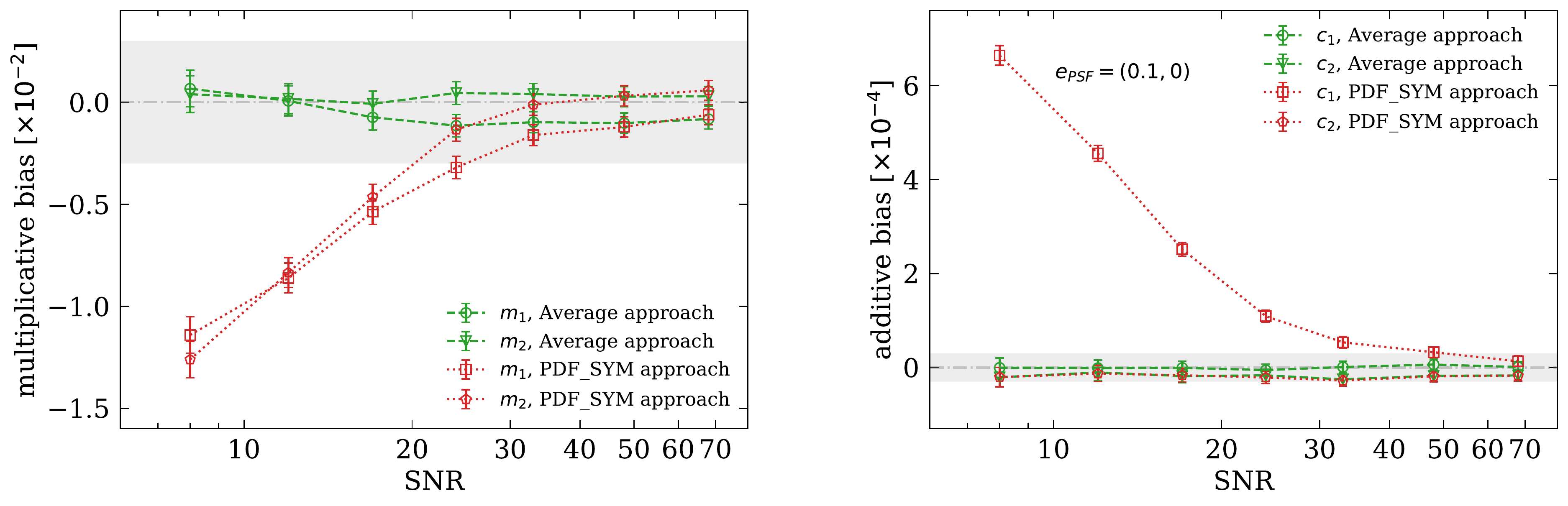}
	\caption{The multiplicative and additive biases measured from the samples of different SNR levels. The shaded region in the left (right) panel, $\pm 3\times 10^{-3}$ ($\pm 3\times10^{-5}$), is the goal to achieve in accuracy. }\label{fig:pts_mc}
\end{figure*}

In the PDF\_SYM approach, we indeed find shear biases for the faint sources. Figure \ref{fig:pts_mc} shows the multiplicative and additive bias measured from the samples of different SNR levels. The multiplicative bias could reach about $1-2\%$ when the SNR $\lesssim 10$, and the PSF anisotropy ($e \sim 0.1$) can introduce an additive bias of order a few times $10^{-4}$ at the similar SNR level. These biases are not seen in the average approach, as shown in the same figure. The results above are measured from the galaxy samples of different SNRs generated by the random walk method. The merit of random walk method includes precise shape distortion, less computational cost, and efficient PSF convolution. (The details of random walk method and image simulation can be found in \cite{Li2020}.)

\begin{figure*}[htbp]
	\centering
	\includegraphics[width=0.9\linewidth]{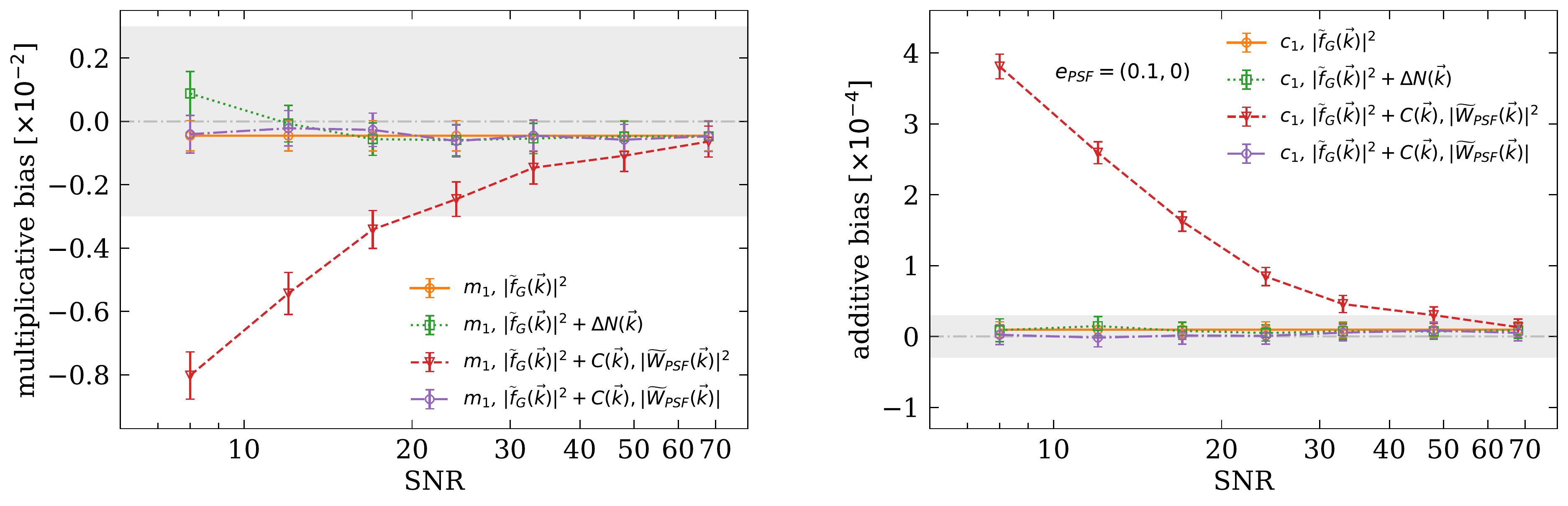}
	\caption{The multiplicative and additive bias at different SNR levels measured by the original PDF\_SYM from different components of the power spectrum of the noisy galaxy image. To avoid the overlap between curves, only the $m_1$'s and $c_1$'s curves are shown. Orange curves: Measured from the power spectrum of the noise-free galaxy image. Green curves: Measured from the combination of noise-free part and the noise power spectrum residual, $\Delta N(\vec{k})$. Red curves: Measured from the combination of noise-free galaxy image power spectrum and the noise-source coupling term $C(\vec{k})$. The PSF power spectrum, $\vert \widetilde{W}_{P}(\vec{k})\vert^2$, is used for deconvolution. Purple curves: Same as the red curves, but with one less power of $\vert\widetilde{W}_{P}(\vec{k})\vert$ in the definition of $T(\vec{k})$ for the coupling term $C(\vec{k})$. The purple curves are for demonstrating the effect mentioned in the last paragraph of \S\ref{sec:bias_cal}.}\label{fig:pts_componets}
\end{figure*}

To reveal the origin of the bias, we should go back to the definition of $M(\vec{k})$ in Eq.(\ref{FQ_TM}). With the existence of noise, the source image is the sum of galaxy and noise image, \ie$f_I(\vec{x}) = f_G(\vec{x}) + f_N(\vec{x})$. We can write $M(\vec{k})$ as: 
\begin{eqnarray}
\label{Eq:Mk}
M(\vec{k}) =  \left\vert\widetilde{f}_G(\vec{k})\right\vert^2 + \Delta N(\vec{k}) + C(\vec{k}).
\end{eqnarray}
The first term is the power spectrum of the noise-free galaxy image. $\Delta N(\vec{k})$ is the residual noise power spectrum after the background noise subtraction defined in Eq.(\ref{FQ_TM}). $C(\vec{k})$ is the coupling between the Fourier transformations of the galaxy and that of the background noise image: 
\begin{eqnarray}
C(\vec{k}) = \widetilde{f}_{G}^{*}(\vec{k})\widetilde{f}_N(\vec{k}) + \widetilde{f}_{G}(\vec{k})\widetilde{f}_{N}^{*}(\vec{k})
\end{eqnarray}
The contributions of $C(\vec{k})$ and $\Delta N(\vec{k})$ to the shear estimators vanish in the ensemble averages defined in Eq.(\ref{method:mean}). Therefore, they do not bias the ensemble average approach as shown in Figure \ref{fig:pts_mc}. However, we recently find that, at the low SNR end, the PDF\_SYM approach can be biased by $C(\vec{k})$, but not by $\Delta N(\vec{k})$. It can be seen in Figure \ref{fig:pts_componets}, in which we show the shear biases when different components are included in the image power spectrum: $\vert\widetilde{f}_G(\vec{k})\vert^2 + \Delta N(\vec{k})$ (green curves) and $\vert\widetilde{f}_G(\vec{k})\vert^2 + C(\vec{k})$ (red curves).  

To understand the bias mechanism, we should consider the contribution of $C(\vec{k})$ to the shear estimators in Eq.(\ref{shear_estimator}) at the low SNR end. In the following part of this section, we calculate the joint PDF of $G_1$ and $G_2$ with the contribution of $C(\vec{k})$ included, and show the asymmetry caused by it in the PDF. For convenience, we neglect the $\Delta N(\vec{k})$ in the calculation since it does not cause the bias. We use $f_k$ and $n_k$ to denote the Fourier transformation of the noise-free galaxy image and that of the noise image respectively. Following Eq.(\ref{shear_estimator})\&(\ref{method:pdf}), the shear estimators to be symmetrized in the PDF\_SYM approach are defined as
\begin{eqnarray}
\label{Eq:gg}
\hat{G}_1&=&\int{d}^2\vec{k} P_k T(\vec{k})\left(\vert f_k\vert^2 + C(\vec{k})\right) \\ \nonumber
&=&\Delta^2\sum_k A_k \left(\vert f_k\vert^2+2f_k^{Re}n_k^{Re}+2f_k^{Im}n_k^{Im}\right), \\ \nonumber
\hat{G}_2&=&\int{d}^2\vec{k} \; Q_k T(\vec{k})\left(\vert f_k\vert^2 + C(\vec{k})\right)\\ \nonumber
&=&\Delta^2\sum_kB_k\left(\vert f_k\vert^2+2f_k^{Re}n_k^{Re}+2f_k^{Im}n_k^{Im}\right),
\end{eqnarray}
where
\begin{eqnarray}
&&A_k = P_k T(\vec{k}),\; B_k = Q_k T(\vec{k}) ,\\ \nonumber
P_k = k_x^2-k_y^2&& - \hat{g}_1\left[2k^2-\beta^2\left(k^4+k_x^4-6k_x^2k_y^2+k_y^4\right)\right],\\ \nonumber
Q_k = 2k_xk_y-&&\hat{g}_2\left[2k^2-\beta^2\left(k^4-k_x^4+6k_x^2k_y^2-k_y^4\right)\right].
\end{eqnarray}
$\hat{g}_{1/2}$ is the assumed value for the shear signal. The integrations are written in discrete forms for technical convenience later. $\Delta$ is simply the pixel size in the Fourier space.

The key for understanding the origin of the bias is to obtain the PDFs of $\hat{G}_1$ and $\hat{G}_2$ in the presence of noise. For this purpose, we can start by writing the PDFs of one galaxy under many different noise realizations. For technical convenience, we need to assume that the noise is mainly the white noise, \ie, the PDFs of $n_k^{Re}$ and $n_k^{Im}$ follow normal distributions, and the standard deviation ($\sigma$) is independent of $\vec{k}$. Then, we can write the joint PDF of $n_k^{Re}$ and $n_k^{Im}$ as:
\begin{equation}
P_N(n_k^{Re},n_k^{Im})=\frac{1}{2\pi\sigma^2}\exp\left[-\frac{\left(n_k^{Re}\right)^2+\left(n_k^{Im}\right)^2}{2\sigma^2}\right].
\end{equation}
Consequently, the joint PDF of $\hat{G}_1$ and $\hat{G}_2$ is given by:
\begin{eqnarray}
\label{Eq:Pgg_1}
&P_G&\left(\hat{G}_1,\hat{G}_2\right)\\ \nonumber
&=&\left\{\prod_k\int\int dn_k^{Re}dn_k^{Im}P_N(n_k^{Re},n_k^{Im})\right\}\\ \nonumber
&&\times\delta_D\left[\hat{G}_1-\Delta^2\sum_KA_K\left(\vert f_K\vert^2+2f_K^{Re}n_K^{Re}+2f_K^{Im}n_K^{Im}\right)\right]\\ \nonumber
&&\times\delta_D\left[\hat{G}_2-\Delta^2\sum_KB_K\left(\vert f_K\vert^2+2f_K^{Re}n_K^{Re}+2f_K^{Im}n_K^{Im}\right)\right]\\ \nonumber
&=&\frac{1}{(2\pi)^2}\left\{\prod_k\int\int dn_k^{Re}dn_k^{Im}P_N(n_k^{Re},n_k^{Im}) \right\}\\ \nonumber
&&\times\int\int dxdy \exp\left(ix\hat{G}_1+iy\hat{G}_2\right)\\ \nonumber
&&\times\exp\left\{-2i\Delta^2\sum_K(xA_K+yB_K)\left(f_K^{Re}n_K^{Re}+f_K^{Im}n_K^{Im}\right)\right.\\ \nonumber
&&\left.-i\Delta^2\sum_K(xA_K+yB_K)\vert f_K\vert^2\right\} 
\end{eqnarray}
The integration limits in the above formula are all from $-\infty$ to $+\infty$, we therefore simply neglect the notations of the limits in the formula. This is the convention we adopt in the rest of the paper as well. To simplify the formula, we can first integrate out the variables $n_k^{Re}$ and $n_k^{Im}$. This can be done by using the following relation:  
\begin{equation}
\int \frac{du}{\sqrt{2 \pi}\sigma} \exp \left(-\frac{u^2}{2\sigma^2}-i\alpha u \right) =\exp \left(-\frac{\sigma^2\alpha^2}{2} \right)
\end{equation}
The results are further integrated over the parameters x and y using the following formula:
\begin{eqnarray}
\int&&\int dx dy \exp \left(ixa+iyb-x^2c-2xyd-y^2e\right)\\ \nonumber
=&&\frac{\pi}{\sqrt{ce-d^2}}\exp\left\{-\frac{cb^2-2abd+ea^2}{4(ec - d^2)}\right\},
\end{eqnarray}
At the end, we arrive at the following results: 
\begin{eqnarray}
\label{Eq:Pgg_1}
&&P_G\left(\hat{G}_1,\hat{G}_2\right)
=\frac{1}{(2\pi)^2}\frac{\pi}{\sqrt{U_1U_3-U_2^2}}\\ \nonumber
&&\times\exp\left[-\frac{\left(\hat{G}_1-\Gamma_1\right)^2U_3+\left(\hat{G}_2-\Gamma_2\right)^2U_1}{4(U_1U_3-U_2^2)}\right.\\ \nonumber 
&&+\left.\frac{2\left(\hat{G}_1-\Gamma_1\right)\left(\hat{G}_2-\Gamma_2\right)U_2}{4(U_1U_3-U_2^2)}
\right].
\end{eqnarray}
where
\begin{eqnarray}
\label{Eq:gamma_u}
\left(\Gamma_1,\Gamma_2\right)&=&\Delta^2\sum_k\vert f_k\vert^2\left(A_k,B_k\right)\\ \nonumber
\left(U_1,U_2,U_3\right)&=&2\sigma^2\Delta^4\sum_k\vert f_k\vert^2\left(A_k^2,A_kB_k,B_k^2\right)
\end{eqnarray}

The PDF\_SYM approach estimates each component of shear independently. Therefore, without loss of generality, let us consider the PDF of $\hat{G}_1$. By integrating out $\hat{G}_2$ in the PDF, we can get:
\begin{eqnarray}
\label{1d_PDF}
P_1(\hat{G}_1)&=&\int_{-\infty}^{\infty}d\hat{G}_2 P_G\left(\hat{G}_1,\hat{G}_2\right)\\ \nonumber
&=&\frac{1}{4\sqrt{\pi  U_1}}\exp\left(-\frac{(\hat{G}_1-\Gamma_1)^2}{4U_1}\right)
\end{eqnarray}
To better understand the implications of Eq.(\ref{1d_PDF}), we can rewrite $\Gamma_1$ and $U_1$ as:
\begin{eqnarray}
\label{1d_PDF_def1}
\Gamma_1&=&\Delta^2\sum_kA_k\vert f_k\vert^2\\ \nonumber
&=&\int{d}^2\vec{k}\;P_k\exp(-\beta^2k^2)\vert \tilde{f}_S(\mathrm{M}^{-1}\vec{k})\vert^2 \\ \nonumber
&=&\int{d}^2\vec{k}\exp(-\beta^2k^2)\vert \tilde{f}_S(\vec{k})\vert^2\\ \nonumber
&&\times\left[k_x^2-k_y^2+4g_2\beta^2k_xk_y(k_x^2-k_y^2)\right]
\end{eqnarray}
\begin{eqnarray}
\label{1d_PDF_def2}
U_1&=&2\sigma^2\Delta^4\sum_kA_k^2\vert f_k\vert^2\\ \nonumber
&=&2\sigma^2\Delta^2\int{d}^2\vec{k} \; P_k^2\exp(-2\beta^2k^2)\frac{\vert \tilde{f}_S(\mathrm{M}^{-1}\vec{k})\vert^2}{\vert \widetilde{W}_{P}(\vec{k})\vert^2}\\ \nonumber
&=&2\sigma^2\Delta^2\int{d}^2\vec{k}\exp(-2\beta^2k^2)\frac{\vert \tilde{f}_S(\vec{k})\vert^2}{\vert \widetilde{W}_{P}(\mathrm{M}\vec{k})\vert^2}\\ \nonumber 
&&\times\left[(k_x^2-k_y^2)^2+8g_2\beta^2k_xk_y(k_x^2-k_y^2)^2\right]
\end{eqnarray}
and
\begin{eqnarray}
\label{1d_PDF_def3}
\mathrm{M}&=&\left[\begin{array}{cc}
1-g_1 &  -g_2 \\
-g_2 &  1+g_1 
\end{array}\right].
\end{eqnarray}
In the above equations, $\tilde{f}_S(\vec{k})$ is the Fourier transformation of the pre-lensing galaxy profile. We have assumed that the $1$-D PDF of $\hat{G}_1$ has been symmetrized by the true shear signals $g_1$ and $g_2$, \ie, $\hat{g}_{1,2}=g_{1,2}$. Note that $P_1(\hat{G}_1)$ only takes into account the contribution from one galaxy (with many different noise realizations). The total PDF of $\hat{G}_1$, called $P_1^{\Sigma}(\hat{G}_1)$, should be the sum of $P_1(\hat{G}_1)$ from many galaxies that are statistically isotropic. In Eq.(\ref{1d_PDF_def1} \& \ref{1d_PDF_def2}), we can see that for every $(\Gamma_1,U_1)$, if there is another galaxy with $(-\Gamma_1,U_1)$, the symmetry of $P_1^{\Sigma}(\hat{G}_1)$ is then guaranteed. For parity reasons, we can study this issue by setting $g_2=0$ in the equations. It is then straightforward to check that the symmetry of the PDF is indeed preserved if the definition of $U_1$ does not contain the factor $1/|\widetilde{W}_{P}(\mathrm{M}\vec{k})|^2$. This additional factor does not only cause multiplicative shear biases (as the elements of matrix $\mathrm{M}$ depends on the shear components), but also additive biases from the anisotropy of the PSF profile $\widetilde{W}_{P}(\vec{k})$. 

The existence of $1/|\widetilde{W}_{P}(\mathrm{M}\vec{k})|^2$ originates from $C(\vec{k})$, \ie, the noise-source coupling term. Since $C(\vec{k})$ only contain the Fourier transformation of galaxy image $\widetilde{f}_{G}(\vec{k})$, but not the power spectrum, the product of $T(\vec{k})C(\vec{k})$ does not cancel out the power spectrum of the PSF, leading to the additional factor $1/|\widetilde{W}_{P}(\mathrm{M}\vec{k})|^2$ in $U_1$. It implies that if we artificially remove the residual PSF power $|\widetilde{W}_{P}(\vec{k})|$ in $T(\vec{k})C(\vec{k})$, the shear bias should vanish. This operation can be realized in simulations. We show in Figure \ref{fig:pts_componets} with the purple curves that the shear biases indeed vanish when the additional PSF power is removed from $T(\vec{k})C(\vec{k})$. Unfortunately, this kind of operation is not feasible in processing real data. Therefore, in practice, we need to consider an alternative way to cancel out the asymmetries in the PDFs of the shear estimators.

\section{Restoring PDF symmetry using additional noise}\label{sec:bias_corr}

\begin{figure*}[htbp]
	\centering
	\includegraphics[width=0.9\linewidth]{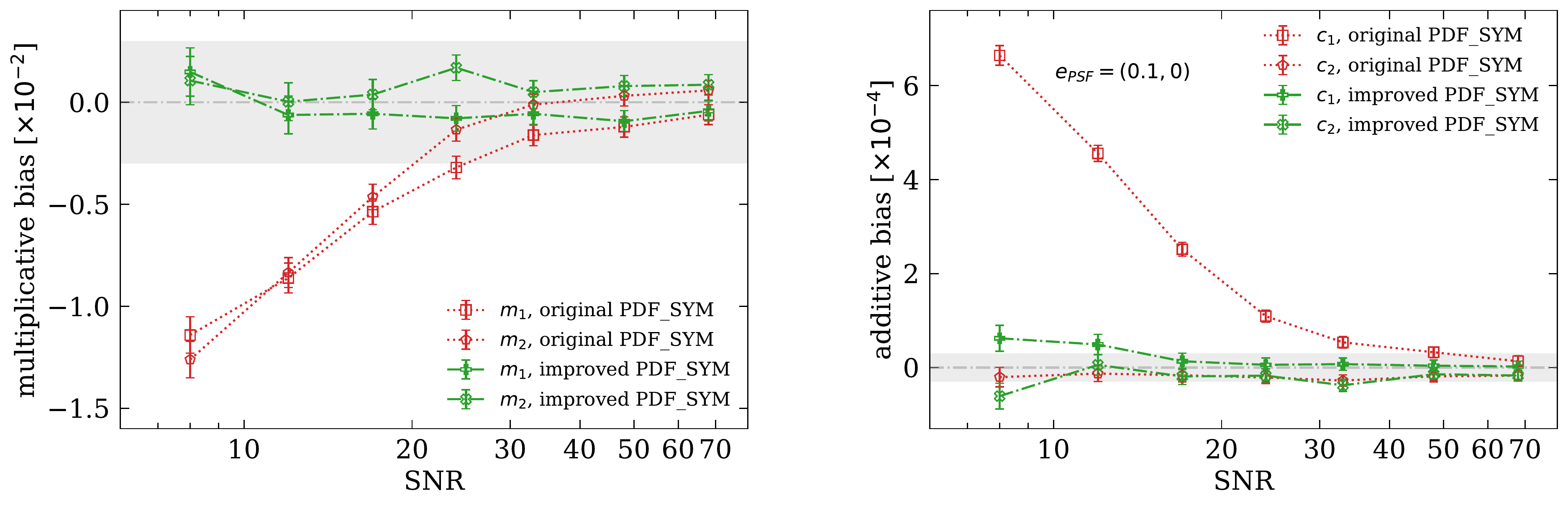}
	\caption{The multiplicative and additive bias measured by the original and improved PDF\_SYM approach from the samples of different SNR respectively. For each SNR level, the multiplicative bias is corrected to smaller than $3\times10^{-3}$ by the improved PDF\_SYM approach. The residual of additive bias is $\sim 5\times10^{-5}$ for the samples with SNR $\le 15$.}\label{fig:pts_mc_corr}
\end{figure*}

Our basic idea is to bring in another noise-source coupling term $C'(\vec{k})$ to compensate for the anisotropy of the PDF caused by the original $C(\vec{k})$. For this purpose, we need another background noise image, say  $f_{B^{\prime}}(\vec{x})$, to couple with the source image through Fourier transformation. To remove the additive bias caused by the PSF anisotropy, we find it useful to rotate the original PSF image by $90$ degrees, and use it in PSF re-convolution for the $C'(\vec{k})$. In detail, building the modified Fourier\_Quad shear estimators includes the following four steps:

\begin{enumerate}
	\item[I.] Obtain the original shear estimators, including $G_1$, $G_2$, $N$, $U$, and $V$, according to definitions in Section \ref{sec:FQ}. This step needs the first noise image, $f_{B}(\vec{x})$, for subtracting the contribution of the background noise.

	\item[II.] Obtain a new noise-source coupling term $C^{\prime}(\vec{k})$ by adding another noise image, $f_{B^{\prime}}(\vec{x})$, to the original source image, $f_I(\vec{x})$. The new source image $f_{I^{\prime}}(\vec{x})$ now contains one galaxy image and two noise realizations as:
	\begin{eqnarray}
	f_{I^{\prime}}(\vec{x}) = f_G(\vec{x}) + f_B(\vec{x}) + f_{B^{\prime}}(\vec{x}).
	\end{eqnarray}
	The new noise-source coupling term $C^{\prime}(\vec{k})$ is then achieved by subtracting the power spectra of the second noise image and the original source image from that of the new image $f_{I^{\prime}}(\vec{x})$, \ie:
	\begin{eqnarray}
	C^{\prime}(\vec{k})&=&\left\vert \widetilde{f}_{I^{\prime}}(\vec{x})\right\vert^2 - \left\vert \widetilde{f}_{I}(\vec{x})\right\vert^2 - \left\vert \widetilde{f}_{B^{\prime}}(\vec{x})\right\vert^2 \\ \nonumber
	 &=& \widetilde{f}_{G}^{*}(\vec{k})\widetilde{f}_{B^{\prime}}(\vec{k}) + \widetilde{f}_{G}(\vec{k})\widetilde{f}_{B^{\prime}}^{*}(\vec{k})\\ \nonumber
	 &+&\widetilde{f}_{B}^{*}(\vec{k})\widetilde{f}_{B^{\prime}}(\vec{k}) + \widetilde{f}_{B}(\vec{k})\widetilde{f}_{B^{\prime}}^{*}(\vec{k}).
	\end{eqnarray}
	Note that the last two terms in the above formula are the noise-noise coupling terms, which do not cause shear biases, as we show previously. We therefore can simply neglect them.
	
	\item[III.] Rotate the PSF image by $90$ degrees\footnote{For a $90$-degree rotation, one simply exchanges the coordinate components. There is no need to perform interpolations on the sub-pixel level.} in the measurement to cancel out the additive bias induced by the anisotropy of PSF. The image rotation changes the factor $T(\vec{k})\rightarrow T'(\vec{k})=\vert \widetilde{W}_\beta(\vec{k})\vert^2 /\vert \widetilde{W}_P^{\perp}(\vec{k})\vert^2$ for $C^{\prime}(\vec{k})$. $\vert \widetilde{W}_P^{\perp}(\vec{k})\vert^2$ is the power spectrum of the rotated PSF image. The corrections to the shear estimators are then defined as:
	\begin{eqnarray}
	\Delta G_1&=&-\int{d}^2\vec{k}\left(k_x^2-k_y^2\right)T'(\vec{k}) C^{\prime}(\vec{k}), \\ \nonumber
	\Delta G_2&=&-\int{d}^2\vec{k}\left(2k_xk_y\right)T'(\vec{k}) C^{\prime}(\vec{k}).
	\end{eqnarray}
	
	\item[IV.] Rotate $\Delta G_1 + i\Delta G_2$ by $45$ degrees, and add them to the original shear estimators $\hG_1 + i\hG_2$ to obtain the corrected shear estimators $\wG_1+i\wG_2$ as: 
	\begin{equation}
	\label{Gnew}
	\left[ \begin{array}{ccc}
	\wG_1 \\
	\wG_2 \\
	\end{array} 
	\right ] = 
		\left[ \begin{array}{ccc}
	\hG_1 \\
	\hG_2 \\
	\end{array} 
	\right ]
	+
	\left[ \begin{array}{ccc}
		0 & -1 \\
		1 & 0\\
		\end{array} 
	\right ]
	\left[ \begin{array}{ccc}
	\Delta G_1 \\
	\Delta G_2 \\
	\end{array} 
	\right ].
	\end{equation}
\end{enumerate}
For shear measurement, one can then take the PDF\_SYM approach to symmetrize the PDF of $\wG_{1/2}$. The improved results are shown in Figure \ref{fig:pts_mc_corr}. The bias is corrected at the cost of a slightly larger error bar ($\sim1.5$ times the orignial one at SNR$\lesssim10$). At the high SNR end, the error bars are similar to the original ones. 

\begin{figure}[htbp]
	\centering
	\includegraphics[width=0.9\linewidth]{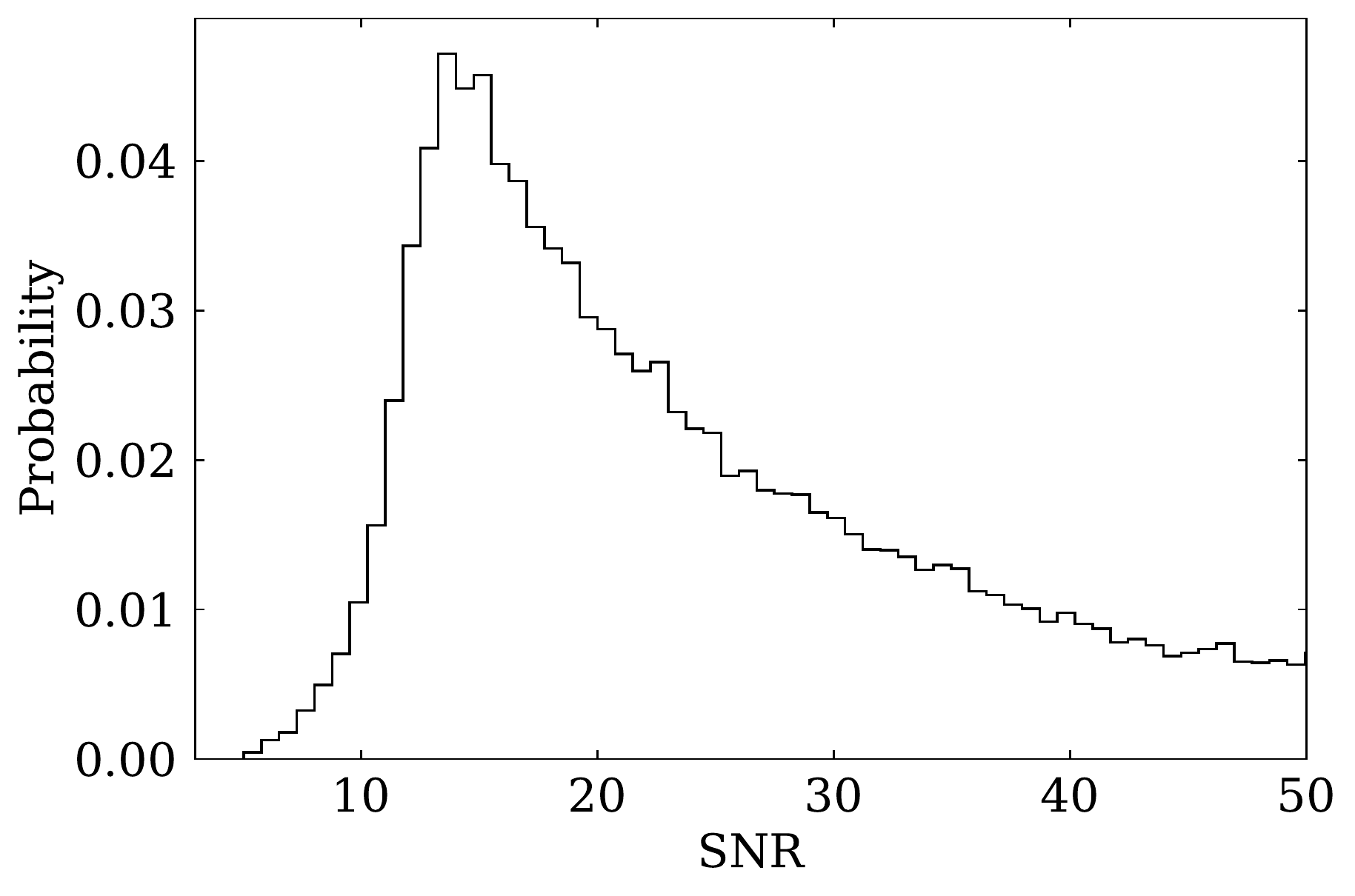}
	\caption{The SNR distribution in the realistic image simulation.}\label{fig:snr_pdf}
\end{figure}

Unlike the single-SNR simulations above, the real survey includes both faint and bright sources. Therefore, we further test the improved PDF\_SYM approach against a more realistic image simulation, in which the magnitudes of galaxies range from $22$ to $25$\footnote{We keep the same setup as the previous image simulations, but assign the fluxes calculated from the magnitudes to the galaxies to mimic the real SNR distribution. We simulate the i-band ($i_{814}$) observations with 600 seconds exposure. The gain and zero point are 1.5 $e$-/ADU and 25.77 $mag$ respectively. The stand deviation of the background noise is 60 ADUs, which is obtained by the least-squares fitting to the CFHTLenS images.}. Figure \ref{fig:snr_pdf} shows the SNR distribution. The results are presented in Table \ref{tb:real_simu}. We can see that the new shear estimators consistently lead to a more accurate shear recovery in the PDF\_SYM approach under the realistic observation condition. Note that even with the original estimators, the shear biases are not as large as those in the previous single-SNR simulations. This is simply because of the existence of bright galaxies in the new sample. 
\begin{table}
\centering
\caption{}\label{tb:real_simu}
\begin{tabular}{ccc}
	\multicolumn{3}{c}{Improvement in a realistic image simulation} \\
	\hline 
	\hline 
	& Multiplicative bias& Additive bias \\ 
	& $10^3m_{1/2}$& $10^4c_{1/2}$ \\
	\hline 
	\multirow{2}{*}{Original PDF\_SYM}&  $-6.08(\pm0.67)$&  $3.26(\pm0.16)$\\ 
	&  $-5.95(\pm0.67)$&  $-0.23(\pm0.16)$\\ 
	\hline 
	\multirow{2}{*}{Improved PDF\_SYM}&  $-0.09(\pm0.74)$&  $-0.11(\pm0.17)$\\ 
	&  $0.77(\pm0.74)$&  $-0.15(\pm0.17)$\\ 
	\hline 
\end{tabular}
\end{table} 

To understand how the bias is corrected by the new terms, let us first consider the joint PDF $P_{\Delta}$ of $(\Delta G_1,\Delta G_2)$. Similar to the calculation of $P_G(\hat{G}_1,\hat{G}_2)$, we have:
\begin{eqnarray}
\label{pdelta}
&&P_{\Delta}\left(\Delta G_1,\Delta G_2\right)=\frac{1}{(2\pi)^2}\frac{\pi}{\sqrt{V_1V_3-V_2^2}}\\ \nonumber
&&\times\exp\left[-\frac{\Delta G_1^2V_3-2\Delta G_1\Delta G_2V_2+\Delta G_2^2V_1}{4(V_1V_3-V_2^2)}\right]
\end{eqnarray}
The parameters $V_1$, $V_2$, and $V_3$ are defined as:
\begin{eqnarray}
\label{1d_PDF_def4}
V_1&=&2\sigma^2\Delta^4\sum_k(\kx2y2)^2T'^2(\vk)\vert f_k\vert^2\\ \nonumber
&=&2\sigma^2\Delta^2\int{d}^2\vk(\kx2y2)^2M^\prime(\vk)\\ \nonumber
V_2&=&2\sigma^2\Delta^2\int{d}^2\vk(2k_xk_y)(\kx2y2)M^\prime(\vk)\\ \nonumber
V_3&=&2\sigma^2\Delta^2\int{d}^2\vk(2k_xk_y)^2M^\prime(\vk)
\end{eqnarray}
and 
\begin{equation}
M^\prime(\vk)=\exp(-2\beta^2k^2)\vert \tilde{f}_S(\mathrm{M}^{-1}\vk)\vert^2\frac{\vert W_{P}(\vk)\vert^2}{\vert W_{P}^{\perp}(\vk)\vert^{4}}
\end{equation}
To compensate for the asymmetry of the PDF, we rotate ($\Delta G_1$,$\Delta G_2$) by $45$ degrees, and then add them to ($\hG_1$,$\hG_2$) respectively, as shown in Eq.(\ref{Gnew}). The joint PDF of ($\wG_1$, $\wG_2$) can be written as a convolution between the $P_G(\hG_1,\hG_2)$ and $P_{\Delta}(\Delta G_1,\Delta G_2)$. Consequently, we can write down the corresponding new $1$-D PDF for $\hG_1$, as:
\begin{eqnarray}
\label{PF}
P_F&&(\wG_1)\\ \nonumber
&&=\int d\wG_2\int d\Delta G_1 \int d\Delta G_2  P_{\Delta}(\Delta G_1,\Delta G_2)\\ \nonumber 
&&\quad \times P_G(\wG_1+\Delta G_2,\wG_2-\Delta G_1)\\ \nonumber
&&=\frac{1}{2\sqrt{\pi(U_1+V_3)}}\exp\left\{-\frac{(\wG_1-\Gamma_1)^2}{4(U_1+V_3)}\right\},
\end{eqnarray}
We give the detailed calculations of Eq.(\ref{PF}) in the Appendix. 

To gain some insights in the term $U_1+V_3$, let us assume that the isophotes of the PSF power spectrum are ellipses with the same ellipticity components $e_1$ and $e_2$. We can then approximate the functional form of $\vert W_{P}(\vk)\vert^2$ with Taylor expansion as:
\begin{equation}
\vert W_{P}(\vk)\vert^2=\phi(k^2)-\phi'\cdot\left[2e_1(\kx2y2)+4e_2k_xk_y\right],
\end{equation}
in which $\phi'=d\phi/d(k^2)$. We then have:
\begin{eqnarray}
&&\vert W_{P}(\mathrm{M}\vk)\vert^2\\ \nonumber
&=&\phi(k^2)-\phi'\cdot\left[2(e_1+g_1)(\kx2y2)+4(e_2+g_2)k_xk_y\right] \\ \nonumber
&&\vert W_{P}^{\perp}(\mathrm{M}\vk)\vert^2\\ \nonumber
&=&\phi(k^2)-\phi'\cdot\left[2(g_1-e_1)(\kx2y2)+4(g_2-e_2)k_xk_y\right] 
\end{eqnarray}
Consequently, we can write $U_1 + V_3$ as:
\begin{eqnarray}
&&U_1 + V_3=2\sigma^2\Delta^2\\ \nonumber
&\times&\int_0^{\infty}k^5dk\int_0^{2\pi}d\theta_k\exp(-2\beta^2k^2)\phi^{-1}(k^2)\vert \tilde{f}_S(k,\theta_k)\vert^2\\ \nonumber
&&\times\left\{1+g_1k^2\left[(\beta^2+2\phi'/\phi)\cos(2\theta_k)-\beta^2\cos(6\theta_k)\right]\right.\\ \nonumber
&&+\left.2e_1k^2\cos(6\theta_k)\phi'/\phi\right\}
\end{eqnarray}
in which we have set $g_2=e_2=0$ for convenience, as they do not affect our discussion of $g_1$. We can see that $U_1 + V_3$ is made of a scalar component $N_s$ and a non-scalar quantity $\delta_{N_s}$ (containing spin-2 and spin-6 components) defined as:
\begin{eqnarray}
&&N_s=2\sigma^2\Delta^2\\ \nonumber
&\times&\int_0^{\infty}k^5dk\int_0^{2\pi}d\theta_k\exp(-2\beta^2k^2)\phi^{-1}(k^2)\vert \tilde{f}_S(k,\theta_k)\vert^2\\ \nonumber
&&\delta_{N_s}=2\sigma^2\Delta^2\\ \nonumber
&\times&\int_0^{\infty}k^7dk\int_0^{2\pi}d\theta_k\exp(-2\beta^2k^2)\phi^{-1}(k^2)\vert \tilde{f}_S(k,\theta_k)\vert^2\\ \nonumber
&\times&\left\{g_1(\beta^2+2\phi'/\phi)\cos(2\theta_k)+(2e_1\phi'/\phi-g_1\beta^2)\cos(6\theta_k)\right\}
\end{eqnarray}
Note that the non-scalar part $\delta_{N_s}$ can be treated as a small perturbation to the scalar component $N_s$. Furthermore, in the 1-D PDF $P_F(\wG_1)$, $\Gamma_1$ can be treated as a quantity much smaller than $\wG_1$, as we are dealing with sources of very low signal-to-noise ratios here\footnote{$\Gamma_1$ and $\wG_1$ are measured from the power spectrum of noise-free image and the noisy image respectively. The amplitude of the latter's power spectrum is much higher than that of the former's in case of very low SNR. Therefore, $\Gamma_1$ is much smaller than $\wG_1$.}. $P_F(\wG_1)$ can therefore be expanded as:
\begin{eqnarray}
&&P_F(\wG_1)\propto\frac{1}{\sqrt{N_s}}\exp\left(-\frac{\wG_1^2+\Gamma_1^2}{4N_s}\right)\\ \nonumber
&\times&\left\{1+\frac{\wG_1\Gamma_1}{2N_s}\left(1-\frac{\delta_{N_s}}{N_s}\right)+\frac{\delta_{N_s}}{N_s}\left(\frac{\wG_1^2+\Gamma_1^2}{4N_s}-\frac{1}{2}\right)\right\}
. \\ \nonumber
\end{eqnarray} 
Neglecting $\Gamma_1^2$, and averaging over an ensemble of galaxies of random orientations, we get:
\begin{equation}
\bar{P}_F(\wG_1)\propto\frac{1}{\sqrt{N_s}}\left(1-\frac{\wG_1}{2N_s^2}\langle\Gamma_1\delta_{N_s}\rangle\right)\exp\left(-\frac{\wG_1^2}{4N_s}\right).
\end{equation} 
There is still an asymmetric part of $\bar{P}_F(\wG_1)$:
\begin{eqnarray}
\label{asym_f}
\bar{P}_F^{ASYM}(\wG_1)\propto\frac{-\wG_1}{2N_s^2\sqrt{N_s}}\langle\Gamma_1\delta_{N_s}\rangle\exp\left(-\frac{\wG_1^2}{4N_s}\right).
\end{eqnarray} 
However, we find that the term $\langle\Gamma_1\delta_{N_s}\rangle$ does not contain contributions from the PSF ellipticity, which now contributes a spin-6 component to $\delta_{N_s}$, therefore does not couple with $\Gamma_1$, a spin-2 quantity. In principle, there should still be some small amount of multiplicative bias on $g_1$, as $\delta_{N_s}$ does contain a spin-2 term from $g_1$. The magnitude of this bias is very small however, as $\Gamma_1$ is already very small comparing to $\wG_1$. This fact can also be seen in the following way: if the PSF in $T'(k)$ is not rotated by 90 degrees, there is some residual additive bias of order $10^{-4}$ when the PSF ellipticity is 0.1, because $\delta_{N_s}$ in this case contains a term of spin-2 from $e_1$. As $g_1$ is an order of magnitude smaller than 0.1, the shear bias introduced by Eq.(\ref{asym_f}) must be of order $10^{-5}$, which is not significant for our purpose.

\section{conclusion}\label{sec:conclusion}
In this paper, we focus on the PDF\_SYM approach in the Fourier\_Quad method and find that it can be slightly biased by the faint sources. The typical magnitude of the multiplicative bias can reach $\sim1\%$ level at SNR lower than $10$. The anisotropy of PSF can cause additive biases that could reach about $6\times10^{-4}$ at the same SNR level when $e_{PSF} $ = $0.1$. Based on the image simulations, we find that the bias is caused by the asymmetry in the PDF of shear estimators, originating from noise-source coupling. The coupling term has been neglected in our previous works since it vanishes in the ensemble average approach, and only biases the PDF\_SYM approach slightly at the very faint end. Furthermore, we show in detail that the imprint of the PSF power spectrum in the coupling term is the underlying cause of both the multiplicative and additive biases.

By introducing slight modifications to the original Fourier\_Quad shear estimators, we find that the multiplicative bias can be suppressed from $1\%$ to less than $3\times10^{-3}$ at SNR smaller than 10, and the additive bias can be corrected to less than $5\times 10^{-5}$. Meanwhile, the correction does not affect the accuracy at the high SNR end. Due to the additional noise introduced to the measurement, we correct the biases at the cost of a slightly larger error bar that is $\sim 1.5$ times the original one at the faint end. However, the improved method allows us to make use of more faint sources, which account for a significant portion of the sample, but are often excluded from shear measurements.

\acknowledgments

The computation resources of this work are provided by the Department of Astronomy and the $\pi$ 2.0 cluster supported by the Center for High Performance Computing at Shanghai Jiao Tong University. This work is supported by the National Key Basic Research and Development Program of China (No.2018YFA0404504), and the NSFC grants (11673016, 11621303, 11890691, 12073017).

\appendix
\label{app}

Following the definitions in Eq.(\ref{Eq:Pgg_1}, \ref{Gnew}, \ref{pdelta}), we can derive the PDF of $\wG_1$ as:
\begin{eqnarray}
&&P_F(\wG_1)\\ \nonumber
&&=\int d\wG_2\int d\Delta G_1\int d\Delta G_2 P_G(\wG_1+\Delta G_2,\wG_2-\Delta G_1)P_{\Delta}(\Delta G_1,\Delta G_2)\\ \nonumber
&&=\frac{1}{16\pi^2}\frac{1}{\sqrt{U_1U_3-U_2^2}\sqrt{V_1V_3-V_2^2}}\int d\Delta G_1\int d\Delta G_2\int d\wG_2\exp\left\{-\frac{V_3\Delta G_1^2+V_1\Delta G_2^2-2V_2\Delta G_1\Delta G_2}{4(V_1V_3-V_2^2)}\right\} \\ \nonumber
&&\times\exp\left\{-\frac{U_3(\wG_1-\Gamma_1+\Delta G_2)^2+U_1(\wG_2-\Gamma_2-\Delta G_1)^2-2U_2(\wG_1-\Gamma_1+\Delta G_2)(\wG_2-\Gamma_2-\Delta G_1)}{4(U_1U_3-U_2^2)}\right\}\\ \nonumber
&&=\frac{\sqrt{\pi}}{8\pi^2}\frac{1}{\sqrt{U_1}\sqrt{V_1V_3-V_2^2}} \int d\Delta G_1\int d\Delta G_2\exp\left\{-\frac{(\wG_1-\Gamma_1+\Delta G_2)^2}{4U_1}-\frac{V_3\Delta G_1^2+V_1\Delta G_2^2-2V_2\Delta G_1\Delta G_2}{4(V_1V_3-V_2^2)}\right\} \\ \nonumber
&&=\frac{1}{4\pi}\frac{1}{\sqrt{U_1}\sqrt{V_3}}\int d\Delta G_2\exp\left\{-\frac{(\wG_1-\Gamma_1+\Delta G_2)^2}{4U_1}-\frac{\Delta G_2^2}{4V_3}\right\} \\ \nonumber
&&=\frac{1}{2\sqrt{\pi(U_1+V_3)}}\exp\left\{-\frac{(\wG_1-\Gamma_1)^2}{4(U_1+V_3)}\right\} \\ \nonumber
\end{eqnarray}

\end{document}